\documentclass{webofc}
\usepackage[varg]{txfonts}   
\usepackage{graphicx}
\usepackage{pdfpages}
\makeatletter
\renewcommand\@biblabel[1]{}
\makeatother
\newgeometry{left=4cm, right=2cm, top=2cm, bottom=2cm}
\begin{document}
\pagestyle{plain}
\pagenumbering{arabic}
\title{Molecular gas dynamics around nuclei of galaxies }
%

\author{\firstname{Francoise} \lastname{Combes}
\inst{1}\fnsep\thanks{\email{francoise.combes@obspm.fr}} 
}

\institute{Observatoire de Paris, LERMA, Coll\`ege de France, CNRS, PSL University, Sorbonne University, Paris, France
          }

\abstract{%
Recent molecular line observations with ALMA in
 several nearby Seyferts have revealed the
existence of molecular tori, and the nature of gas flows
at 10-20~pc scale.
 At 100~pc scale, or kpc-scale, previous NOEMA work on gravitational torques
had shown that only about one third of Seyfert galaxies experienced
molecular inflow and central fueling, while in most cases the
gas was stalled in rings.
At higher resolution, i.e. 10-20~pc scale, it is possible
now to see in some cases AGN fueling due to nuclear trailing spirals,
influenced by the black hole potential. This brings
smoking gun evidence for nuclear fueling.
 In our sample galaxies, the angular resolution of up to 60~mas
 allows us to reach the black hole (BH) sphere of influence and the BH mass
can be derived more directly than with the M-sigma relation.
	}
\maketitle
\section{Introduction}
\label{intro}
Our vision of the close environment of the black hole and its accretion disk
has evolved significantly in the recent years. While the geometrical orientation
paradigm helped to understand the difference between type 1 and 2 AGN, through
the presence of a dusty torus at parsec scale (e.g. Urry \& Padovani 1995), 
it is possible that the required obscuration is only due to outflowing dust
in a hollow polar cone (Asmus et al. 2016). Around the accretion disk,
and the ionised Broad Line Region (BLR), dust cannot survive due to high
temperatures, but at the sublimation radius, radiation pressure pushes
the dust in the polar cone.  Gas and dust are inflowing through a 
molecular disk, where sometimes water masers are seen,
and also H$_2$ and CO emission lines. The disk is likely to be flaring,
i.e. its thickness is increasing strongly with radius (H\"onig, 2019).

In the following, I will describe:

1- how the AGN feeding occurs, how the gas angular momentum is transferred 
through gravity torques from
dynamical features, nuclear bars and spirals, and how the gas
accumulates in a molecular torus

2- how the AGN feedback develops, aided by supernovae feedback. 
The two main modes, radiative or kinetic mode, through
winds and radio jets, may occur simultaneously, and result in
significant molecular outflows at intermediate scales.

\section{Fueling}
\label{sec-1}

\begin{figure}[h]
\centering
\includegraphics[width=13cm,clip]{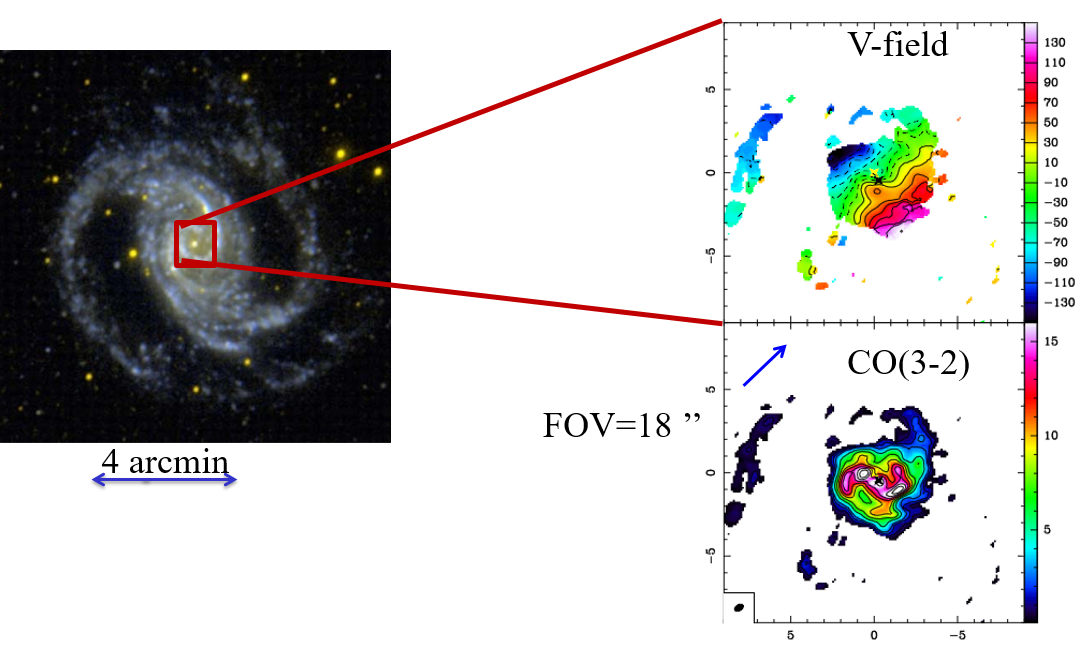}
\caption{NGC 1566 observed in CO(3-2) with ALMA. The left image shows
the large-scale barred spiral structure, and the inner Lindblad resonance (ILR) ring
inside the bar, corresponding to the FOV of 18''. The two first moments at
right reveals a trailing nuclear spiral, inside a nuclear disk, with relatively regular velocity field (adapted from Combes et al. 2014). }
\label{fig-1}       
\end{figure}

One of the first galaxy mapped with ALMA at high angular
resolution is NGC 1566 (Combes et al. 2014).
This is a barred spiral galaxy, with an inner Lindblad resonance ring 
inside the bar. The field of view (FOV) of ALMA at the CO(3-2) line at 345.796 GHz
is 18'', ending just at the border of the ring,
which has then a low signal-to-noise ratio..
Inside the resonant ring, there exists a nuclear disk,
revealed by ALMA. This disk shows a nuclear spiral, which is trailing,
i.e. the same winding sense as the spiral structure outside of the bar,
as seen in figure~\ref{fig-1}.

\subsection{Gravity torques}
\label{sec-11}
The main issue in the fueling process is to get rid of the
high angular momentum of the gas. While viscous torques are
inefficient in galactic disks down to sub-pc scales, gravity torques
from tidal forces in the outer parts, and bars in the inner
parts, are able to exchange angular momentum in between resonances
(e.g. Buta \& Combes 1996). Bars are ending just before their corotation (CR) radius,
where the bar pattern speed equals the angular speed of the tars and gas.
Then the torques exerted on spiral arms just outside the bar are positive, and
drive the gas to the OLR (Outer Lindblad Resonance). Inside the CR radius, torques
are negative down to the ILR, but then change sign inside ILR (Inner Lindblad resonance),
and may give rise to a leading spiral. This ensures that gas is 
accumulating at the ILR ring. 

\begin{figure}[h]
\centering
\includegraphics[width=13cm,clip]{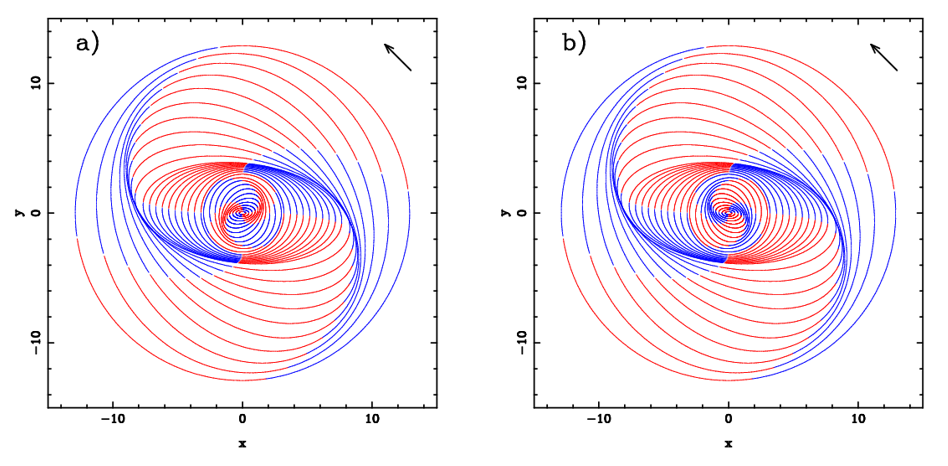}
\caption{
{\bf Left:} Schematic representation of the gas streamlines in a barred galaxy, where ILRs exist. 
The bar is horizontal. When there is no central mass concentration, and the precessing rate 
$\Omega-\kappa/2$ decreases towards the center, the gas is crowding in a leading nuclear spiral, 
and the torque exerted is positive. 
{\bf Right:} When there is a supermassive black hole dominating inside the ILR, the precessing 
rate $\Omega-\kappa/2$ increases again towards the center, and the dissipating gas follows 
now a trailing spiral. The torque is negative, and the gas may fuel the nucleus.
The arrow indicates the sense of rotation. The colors indicate the sign of the torques,
blue: negative and red: positive.}
\label{fig-cartoon}       
\end{figure}

However, the precessing rate $\Omega-\kappa/2$ of the elliptical orbits,
which is usually an increasing function of radius inside ILR,
may change behaviour close to the nucleus, due to the gravitational
influence of the black hole (BH). The rotational velocity $\Omega$,
the epîcyclic frequency $\kappa$, and the precessing rate
$\Omega-\kappa/2$ are then dominated by the keplerian 
potential, and decrease with radius in the central 10-20~pc.
This reverses the winding sense of the gas spiral, and trailing
nuclear spirals are expected in the sphere of influence (SoI) of the BH
cf Figure \ref{fig-cartoon}.
This also reverses the sign of torques, which are now negative,
and drive the gas to the center, fueling the BH.

\subsection{Trailing nuclear spirals, molecular tori}
\label{sec-12}

\begin{figure*}
\centering
\includegraphics[width=12cm,clip]{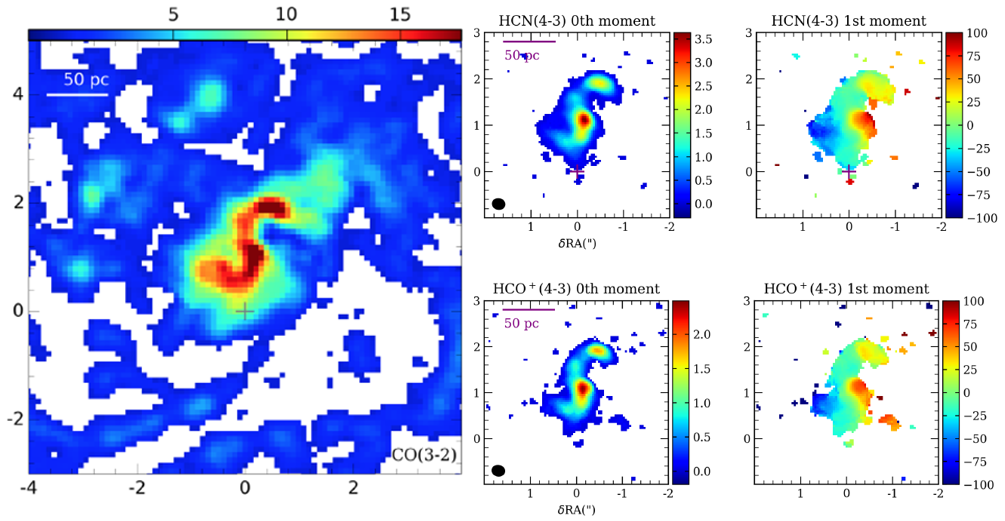}
\caption{Nuclear spiral detected in NGC~1808 in CO(3-2) (left) and
HCN(4-3), HCO+(4-3) (right), with a beam of 0.08'' = 4~pc. 
The velocity field inside the
nuclear spiral has a different major axis orientation from the large-scale one  (adapted from Audibert et al. 2021).}
\label{fig-2}       
\end{figure*}

The trailing nuclear spiral detected in NGC 1566 explains the feeding 
of the AGN. These nuclear spirals are frequently observed, when
the angular resolution is sufficient. It is also detected in NGC 613
with 0.09’’$\times$0.06’’ resolution (5~pc, Combes et al. 2019, Audibert et al. 2019)
but not with 0.5'' resolution (Miyamoto et al. 2017).
It is detected in NGC 1808 (Audibert et al. 2021), and in dense
gas tracers HCO+, HCN and CS, as seen in  
 figure~\ref{fig-2}.

Inside these nuclear spirals, a molecular disk is detected, with apparent
decoupled morphology and kinematics. We call these nuclear disks molecular tori,
they are detected in most of our observed sample of Seyferts 
(Combes et al. 2019). They might be related to the dusty tori of
the classical AGN unification scheme. Their orientation appears randomly distributed with
respect to that of the large-scale galaxy disk. The sizes of molecular
tori are typically 10-20~pc, and are contained in the SoI. Their molecular
mass is of the order of a few 10$^7$ M$_\odot$. 

\begin{figure}
\centering
\includegraphics[width=6cm,clip]{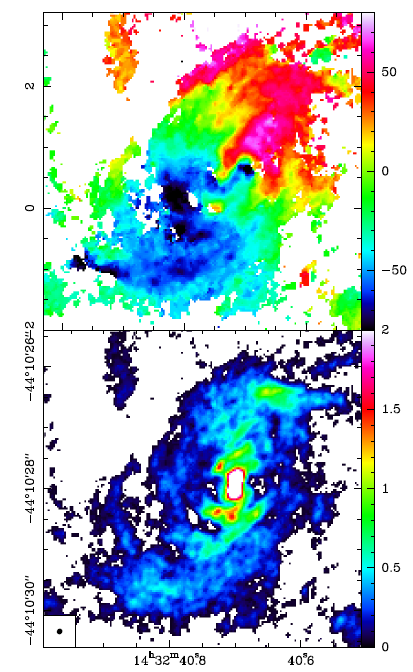}
\includegraphics[width=6cm,clip]{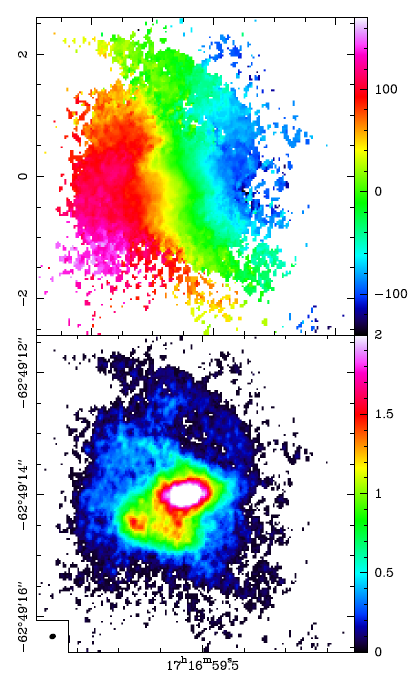}
\caption{Two galaxies, NGC 5643 (left) and NGC 6300 (right) observed
in CO(3-2) with ALMA, from the GATOS sample (Garcia-Burillo et al. 2021).
The two first moments show the more or less tilted kinematical major axis
in the nucleus (adapted from Poitevineau et al. 2022).}
\label{fig-3}       
\end{figure}

When the resolution is enough inside the BH sphere of influence, it is possible
to determine the BH mass, with a proper modelisation of the tilted
molecular torus (e.g. Poitevineau et al. 2022).  Figure~\ref{fig-3} shows
examples of galaxies from the GATOS survey (Garcia-Burillo et al. 2021),
where this is possible, albeit the tilted orientation of the nuclear
disk.

\section{Feedback}
\label{sec-2}

Molecular outflows are frequently observed in nearby Seyferts, and can
be quite small ($<$1kpc), with low velocity $\sim$ 100~km/s (NGC 1433, 
Combes et al. 2013). They are often 
associated to small radio jets (NGC 613, Audibert el al. 2019), although 
sometimes very collimated molecular outflows can occur without a detectable 
radio jet (NGC 1377, Aalto et al. 2016, 2019).

Several mechanisms have been invoked to produce these molecular outflows,
some are purely due to star formation feedback, but then their power is not
related to the AGN luminosity (Cicone et al. 2014). If the AGN is more
luminous than 0.01 Eddington luminosity, radiation pressure on the ionised gas
triggers a fast outflow in the nucleus, which then entrains the molecular gas.
In the case of low-luminosity AGN (lower than 0.01 Eddington), only radio
jets can entrain the molecular gas (kinetic mode).
 At the transition luminosities, both can be present, in particular 
 radio jets and winds driven by radiation pressure on dust
 (Sadowski et al. 2013).

With high angular resolution, it is possible to disentangle AGN from 
 supernovae feedback: for instance, the galactic wind in NGC 1808,
 must be a starburst driven wind, since there is no molecular outflow 
 towards the nucleus (Audibert et al. 2021).
 
 Radio jets appear quite coupled to galaxy disk, to produce
 molecular outflows, since they are not in general perpendicular to the disk. 
 Frequently they can sweep out gas from the inner disk, as in NGC~1068
 where the outflow rate reaches 63 M$_\odot$/yr 
 (Garcia-Burillo et al. 2014, 2016). This prototypical Seyfert-2 reveals
 an edge-on  molecular torus, perpendicular to the radio jet, and a molecular
 outflow, plus a hollow polar cone seen beautifully in polarised 
 near-infrared light
 (Gratadour et al. 2015). The column density in the molecular torus
 and extended dust cone is sufficient to account for the BLR obscuration.

 The most collimated molecular outflow has been found in the early-type
 galaxy NGC 1377 (Aalto et al. 2016, 2019). The molecular jet is almost in
 the plane of the sky, precessing by about 10$^\circ$, and therefore
 it alternates in radius blue and red-shifts. While a radio source is detected
 in the center at cm wavelengths, there is no radio jet within the detection limit,
 which might yield insight in the jet life-time. The relativistic jet could
 disappear in less than 3000 yr, while the molecular jet will remain
 for more than 3 Myr.

\section{Summary}
\label{sec-3}

High angular resolution  has brought a lot of informations about the
AGN fueling. Non-axisymmetric features
(bars, spirals, interactions)  produce gravity torques, allowing
exchange of angular momentum. The gas may be quicky driven towards the center,
through a trailing nuclear spiral, and 
accumulate in a molecular torus,
with decoupled morphology and kinematics.

High resolution can also help disentangle the origin
of the feedback, AGN or supernovae. Molecular outflows are frequent
and may be driven by radio jets or winds (or both).
In case of AGN feedback, the molecular outflows are misaligned with
the rest of the large-scale disk. The outflows are more likely
energy-driven with AGN, while momentum-driven with starbursts.
In the latter case, the feedback is weaker, with the ejected gas
rapidly infalling back after a fountain episode, only
postponing star formation without quenching it.

%
%

\end{document}